# Feature-Gathering Dependency-Based Software Clustering Using Dedication and Modularity

Kenichi Kobayashi, Manabu Kamimura, Koki Kato, Keisuke Yano, Akihiko Matsuo
Software Systems Laboratories, Fujitsu Laboratories Ltd.
Kawasaki, Japan
{kenichi, kamimura.manabu, kato_koki, yano, a_matsuo}@jp.fujitsu.com

*Abstract*—Software clustering is one of the important techniques to comprehend software systems. However, presented techniques to date require human interactions to refine clustering results. In this paper, we proposed a novel dependency-based software clustering algorithm, *SArF*. SArF has two characteristics. First, SArF eliminates the need of the omnipresent-module-removing step which requires human interactions. Second, the objective of SArF is to gather relevant software features or functionalities into a cluster. To achieve them, we defined the *Dedication* score to infer the importance of dependencies and utilized Modularity Maximization to cluster weighted directed graphs. Two case studies and extensive comparative evaluations using open source and industrial systems show that SArF could successfully decompose the systems fitting to the authoritative decompositions from a feature viewpoint without any tailored setups and that SArF was superior to existing dependency-based software clustering studies. Besides, the case studies show that there exist measurable authoritativeness limits and that SArF nearly reached the limits.

## I. INTRODUCTION

Understanding the architecture of a software system is one of the important steps in software maintenance, because the architectural knowledge is often lost or outdated [1]. Software clustering is a technique which decomposes a given system into several subsystems or groups of modules (source files, classes, or other software entities) with manageable sizes. Such decomposition can be used as the architectural knowledge and the high-level abstraction views of the system [2].

Software clustering works well not only for systems written in legacy languages without sophisticated package system, but for modern systems. Through our industrial experience and observations, we found the package structures of modern enterprise systems tend to be aligned to the frameworks such as the three-tier architecture. In such cases, the package structures of the systems provide little additional architectural information beyond the framework. Various software clustering approaches exist, and each of them has its own objective and provides architectural views according to its objective. For example, to find modular subsystems, clustering algorithms finding high-cohesion and low-coupling decomposition [2] is appropriate. Thus, the developers can choose a software clustering technique that is suitable for their objective.

In this paper, we propose a novel dependency-based software clustering algorithm, SArF. The objective of SArF is to gather relevant features into a cluster. Although the term *feature* has many meanings in various contexts, we used the definition in the feature location studies: A (software) feature is a functionality of the system that can be triggered by an external user [3]. We emphasize that SArF does not locate features. SArF only directs relevant features into the same cluster. However, SArF works only using static dependency information, which can be easily collected.

We focus on software clustering using the dependency information of a system. Although existing approaches [2][4][5][6] are semi-automated or automated, refinement feedback processes with human interactions are still required to reach satisfactory final outcomes [1][7]. One of the important issues which require human interactions is the existence of omnipresent modules [8]. Omnipresent modules are modules which connect to several parts of a system but do not seem to belong to any particular subsystem [2]. Since omnipresent modules look like noises in dependency information [8], many studies suggested that removing them would be useful to refine clustering results [2][9][10][11]. Removing omnipresent modules can be automated [4][8][12]; however, the decision whether the omnipresent-module-removing step is used or not and the parameters of the step should be made by a human, and the validity of the removed modules has to be checked manually. If the omnipresent-module-removing step is not needed, software clustering can be further automated.

SArF is designed to be tolerant of omnipresent modules, and it eliminates the need of the omnipresent-module-removing step. Therefore, the software clustering process using SArF is further automated. The tolerance of omnipresent modules and the feature-gathering nature is the key characteristics of SArF. To achieve them, we defined the *Dedication* score which represents the importance of a dependency relationship based on fan-in analysis, and we utilized Modularity Maximization [13] in the community detection literature to cluster directed graphs weighted by the Dedication score.

The remainder of this paper is organized as follows: we will show related work in section II. Our new software clustering algorithm, SArF will be explained in section III. In section IV, we will describe our experiment design. In section V and VI, we will show the two case studies and the comparative evaluations and will discuss about them. The threats to validity will be discussed in section VII. Finally, we will conclude in section VIII.

## II. RELATED WORK

### A. Software Clustering Algorithms

There have already been various software clustering approaches in the literature. From a standpoint of input information, dependency and structural information is frequently used. Bunch [2] is a graph clustering approach that optimizes the objective function, and its objective is to find high-cohesion and low-coupling modular clusters. ACDC [4] is pattern-based approach that utilizes several rules to cluster modules, e.g., clustering a dominator and its dominated modules. To cluster modules with relevant names, naming conventions are used [14]. Semantic information such as identifiers and comments in source code are also used for more sophisticated natural language processing techniques [15][16]. Statement-level dependency is also used [17]. To understand behavioral properties of a system, dynamic information such as execution traces is used for recovering the architecture [18] and for clustering [11].

From a standpoint of methodologies, hierarchical clustering such as single linkage and complete linkage [9][16][19], non-hierarchical clustering such as k-means [15], pattern matching [4], and graph clustering have been used [2][5][7][20]. Four graph clustering techniques were compared by Bittencourt [20].

In recent years, graph clustering has been rapidly developed in the community detection field especially in biology and social network analysis applications. The Girvan-Newman (GN) algorithm [21] is a top-down graph clustering approach which cuts edges with high edge betweenness measures. GN was evaluated in [5][20] and showed good performance results in small software systems but showed poorer results in larger systems. Modularity Maximization such as the Newman algorithm [13] is a bottom-up approach which merges nodes or clusters. It was evaluated in [6] and showed good performance results; however, its performance also becomes poorer in larger systems as shown in Section V and VI. We utilized Modularity Maximization combined with the Dedication score to achieve better performance, the tolerance of omnipresent modules and a feature-gathering characteristic.

From a standpoint of objectives, a noteworthy approach is proposed by Scanniello et al. [15] which has a definite objective that it detects the layer structure of the software architecture of a system. It first decomposes a system *horizontally* (according to the detected layers) and then decomposes each layer *vertically* (using semantic information).

### B. Evaluation of Software Clustering

To evaluate software clustering algorithms, a commonly used criterion is authoritativeness, i.e., distance or similarity between the decomposition computed by the algorithm and the authoritative decomposition which is manually created by experts of a target system. MoJo [22] counts the minimum *move* and *join* operations required to conform two decompositions. Since MoJo has several defects, MoJoFM was proposed as its refinement [23]. To evaluate decompositions with hierarchical nature, UpMoJo [24] was proposed, which counts *up* operations in addition to *move* and *join*. To take account of relationships between modules, measures considering edges such as EdgeSim [25] have been proposed.

Since there are multiple viewpoints in a system, multiple correct decompositions can coexist. Therefore, even if algorithms poorly fit to some authoritative decompositions, it possibly fit to other authoritative decompositions in the same system. Shtern and Tzerpos [26] pointed out that different clustering algorithms cannot be compared if their objectives are different. They proposed an evaluation framework which uses multiple measures [27].

Wu et al. [19] compared several clustering studies using three criteria, authoritativeness, non-extremity of cluster distribution (NED), and stability. The aims of the criteria are to show some desirable properties for good software clustering. NED is used to check bad extreme situations, i.e., too few large or too many small clusters. Stability is the difference between two outputs of consecutive versions of the same software system. Good algorithms should be stable enough to produce similar clusters when small changes happen. NED and stability can be measured without authoritative decompositions. Most recent studies [6][15][16][20] employed this evaluation framework using relative versions of the criteria.

## III. PROPOSED METHOD

In this section, we explain the SArF algorithm. The algorithm comprises of two main ideas, defining the Dedication score to weight dependency edges and utilizing Modularity Maximization to cluster weighted directed graphs.

### A. Dedication Score

Since omnipresent modules behave as noises in software clustering [8], many studies suggested to remove omnipresent modules [4][12]. However, we assumed scoring each dependency between modules is a better way than removing it, because removal causes loss of information and requires human decision to determine various thresholds and parameters.

We found another motivation in the summarizing execution traces literature. Hamou-Lhadj and Lethbridge reported removing implementation details such as utilities can reveal the summarized views of execution traces [28]. They used fan-in analysis to judge whether a trace entry is removed or not. Summarized traces are potential clues to find features. Patel et al. [11] proposed a software clustering algorithm using such extracted features on the basis of their observation that features constitute a natural grouping of the modules that implement them.

Inspired by both studies, although locating a feature is difficult without dynamic or semantic information, we assumed that a set of features can be gathered into a group of modules that implement them by using the dependency information surrounding the modules, if some importance of dependencies is appropriately scored. We also assumed that such scoring can also be used to score dependencies for omnipresent modules. We defined such a score, named *Dedication* as described below.

From a feature viewpoint, when module A depends on module B, we assumed that the importance of the dependency from module A to module B can be interpreted as how likely module B is dedicated to module A. If module B is dedicated only to module A, module B probably shares the same feature with module A. On the other hand, when many modules depend on module B, module B seems not to be dedicated to some specific modules and does not likely share any feature. We designed the definition based on the consideration. Formally, a dependency graph is denoted as G

= <V, E>, where V is a set of vertices, and E is a set of directed edges. A vertex represents a module (such as a source file, class, and method) or a data entity. A directed edge represents a dependency between two modules, the edge from vertex *A* to vertex *B* is denoted as (*A*, *B*). In a simple case, the definition of the Dedication score D(*A*, *B*) of the edge (*A*, *B*) is as follows:

$$D(A,B) = \frac{1}{\text{fanin}(B)}, \quad (1)$$

where fanin(*B*) is the number of incoming edges to vertex *B*.

In practical situations, hierarchical relationships exist in software entities such as a class and methods. To comprehend a system, methods are often too detailed, and classes are rather manageable. However, if a method-level dependency graph is available, its information is worth considering. The Dedication score $D_M(A,B)$ in a multi-level case is defined as follows:

$$D_M(A,B) = \sum_{m \in M_{AB}} \frac{1}{\text{xm}_{\text{fanin}}(m) \cdot \text{mx}(B)}, \quad (2)$$

where $M_{AB}$ is the set of *B*'s members depended by some of *A*'s members, and $\text{xm}_{\text{fanin}}(m)$ is the number of members outside *B* which depend on member *m*, and mx(*B*) is the number of *B*'s members depended by some external members. Dedication scores in a class-level graph are calculated using its corresponding member-level graph and (2). An example of (2) is shown in Fig. 1, and the figure represents class X is mainly dedicated to class A.

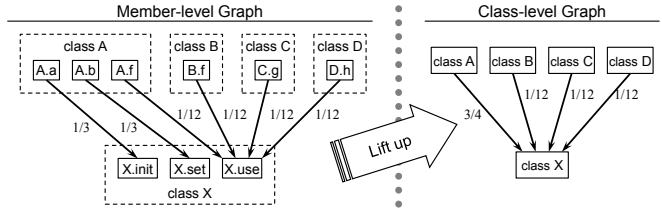

Figure 1. Example of Dedication score calculation (multi-level case)

### B. Modularity Maximization

Since we designed Dedication as the likelihood that source and target modules share common features, the meaningfulness of the weight of edges should be considered. We chose the following criterion of meaningfulness:

$$w - \langle w \rangle, \quad (3)$$

where *w* is the weight of an edge, and ⟨*w*⟩ means the expectation of *w*. If (3) is positive, the edge can be considered meaningful. We utilized a clustering algorithm the strategy of which maximizes the meaningfulness of the weight of all intra-cluster edges.

Modularity Maximization proposed by Newman [13] is one of approaches in community detection and is a bottom-up graph clustering approach which merges leaf nodes or clusters into a larger cluster to maximize the objective function, Modularity Q. Optimizing Modularity Q is NP-hard [29]; however, a greedy search algorithm such as the Newman algorithm [13] can provide a good approximation. The algorithm was enhanced by Clauset et al. [30] and became very fast, and its computational complexity is typically only $O(|V|\log^2|V|)$, where |V| is the number of vertices. Modularity Q can be straightforwardly extended for weighted graphs, and recently extended for directed graphs as follows [31]:

$$Q_D = \frac{1}{W} \sum_{i,j} \left[ A_{ij} - \frac{k_i^{OUT} k_j^{IN}}{W} \right] \delta(c_i, c_j), \quad (4)$$

where *W* is the sum of weight of directed edges in a graph, and $A_{ij}$ is the element of adjacency matrix of the graph or the weight of edge (*i*, *j*), and $k_i^{OUT}$ is the sum of the weight of outgoing edges from vertex *i*, and $k_j^{IN}$ is the sum of the weight of incoming edges to vertex *j*. and $c_x$ is the cluster where vertex *x* belongs, and $\delta(c_i, c_j)$ is Kronecker delta function which equals to 1 if $c_i=c_j$ and 0 otherwise. In (4), the term ($k_i^{OUT} k_j^{IN} / W$) means the expectation of the weight of edge (*i*, *j*), and the term $A_{ij}$ is its actual weight. Thus, (4) matches the criterion (3). Modularity $Q_D$ is interpreted as the magnitude of how much denser the intra-cluster edges are than their expectation. The range of Modularity $Q_D$ is from -1 to 1, and a higher value means a better clustering result.

We used the Newman algorithm with Modularity $Q_D$ in (4) to find clusters from a dependency graph with Dedication. The procedure of the Newman algorithm is briefly explained as follows. The detail is found in [13] [30].

1. Initially, each cluster contains one individual vertex.

2. Find two clusters with the greatest gain or the least loss of $Q_D$ of the merger, and merge them into one cluster. Repeat this step until all clusters are merged. The obtained merge tree (dendrogram) is a hierarchical decomposition.

3. The optimal flat decomposition is the decomposition with the maximum $Q_D$. To obtain it, divide each cluster into its children recursively from the root cluster, while the gain of $Q_D$ of the division is not negative.

### C. SArF Algorithm

We proposed a new software clustering method using the Dedication score and Modularity Maximization. We named the algorithm SArF (Software Architecture Finder). Since the algorithm is deterministic, it has high stability.

The procedure of the algorithm is as follows.

1. Extracting dependency information between modules (classes, members, or other entities) from a target software system.

2. Creating a dependency graph using the extracted information. Member-level information is preferable but not necessary.

3. Calculating the Dedication scores of the dependency graph as described in Section III.A. If the graph is at member-level, the graph is lifted up to the class-level graph.

4. Clustering the graph using directed weighted Modularity Maximization as described in Section III.B.

## IV. EXPERIMENT DESIGN

In this section, we describe the performance measures and the evaluation procedure used in Section V and VI.

### A. Performance Measures

We used authoritativeness in case studies in Section V. In section VI, we additionally used non-extremity of cluster distribution (NED) and stability as in previous studies [6][15][16][19][20].

To evaluate authoritativeness, the authoritative decomposition of a target software system must be obtained to compare the difference between the authoritative decomposition and the decomposition computed by the algorithm. However, obtaining authoritative decomposition is a difficult task, because there are multiple correct decompositions for various viewpoints. Another reason is simple: creating an authoritative decomposition requires heavy efforts. Most of the previous studies used package/directory hierarchies of target systems as their authoritative decomposition. We also used package hierarchies. Additionally, for the case study two, we prepared a manually created authoritative decomposition.

As in the previous studies, we used the following procedure to create an authoritative decomposition from the package hierarchy: First, each cluster corresponds to a package. Then, while a cluster with a size less than or equal to five exists, it is merged into its parent cluster.

To evaluate authoritativeness, we used MoJoSim and MoJoFM [23] measures. MoJoSim (originally called MoJoQ [22]) is the normalized version of MoJo [22]. MoJo is a distance measure between two decompositions defined as MoJo($C,A$) = min(mno($C,A$), mno($A,C$)), where $C$ is the computed decomposition, and $A$ is the authoritative decomposition, and mno($X,Y$) is the minimum required number of *move* and *join* operations of modules to transform decomposition $X$ to decomposition $Y$. MoJoSim and MoJoFM are defined as follows:

$$\text{MoJoSim}(C,A) = \left(1 - \frac{\text{MoJo}(C,A)}{N}\right) \times 100\%,$$
$$\text{MoJoFM}(C,A) = \left(1 - \frac{\text{mno}(C,A)}{N_{maxops}}\right) \times 100\%$$
(5)

where $N$ is the number of modules, and $N_{maxops}$ is the maximum possible value defined as max(mno($\forall X, A$)). For both MoJoSim and MoJoFM, the higher value means $C$ fits to $A$ better, and the value of 100% means $C$ is equal to $A$. Two simple examples are illustrated in Fig. 2.

MoJo has a flaw that a computed decomposition with a few large clusters is overestimated [23]. For example, in Fig. 2, decomposition C is apparently more useful than decomposition D, because dividing a cluster is more difficult than merging two clusters for humans. In the examples, MoJoSim goes against the expectation. Since MoJoFM is more reasonable, we used it where possible. However, since majority of previous studies used MoJoSim, we also measured MoJoSim values.

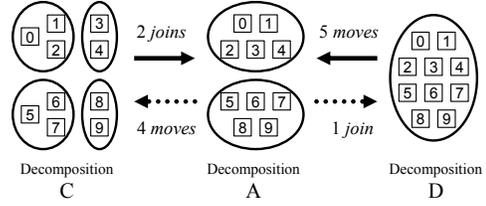

| MoJoSim(C,A) = 1 − min(2,4)/10 = 80% | MoJoSim(D,A) = 1 − min(5,1)/10 = 90% |
|---|---|
| MoJoFM(C,A) = 1 − 2 / 8 = 75% | MoJoFM(D,A) = 1 − 5 / 8 = 37.5% |

Figure 2. Examples of MoJoSim and MoJoFM

NED is defined as follows:

$$\text{NED}(C) = \frac{1}{N} \sum_{c \in C; 5 \leq |c| \leq \max(20, N/5)} |c|, \quad (6)$$

where $c$ is a cluster in $C$, and $|c|$ is the cardinality of $c$.

Stability is defined as Stability($C_i$) = MoJoSim($C_{i-1}$, $C_i$), where $C_i$ is the decomposition of the i-th version of a target system, and $C_{i-1}$ is the decomposition of the previous version. For the purpose of measuring stability, the bi-directional nature of MoJoSim is appropriate rather than unidirectional MoJoFM.

*B. Evaluation Procedures*

The evaluation procedure in this paper is shown in Fig. 3. All target software systems used in this paper are written in Java. We used only jar files as input. First, the jar files of the target software system are collected. Then, at the extracting step, the member-level and class-level dependency graphs are extracted from the jar files using a byte code analyzer based on Javassist (http://www.javassist.org/). Extracted types of dependencies are method invocation, field read, field write, inheritance, and class type reference. To express a class type reference in the member level graph, a virtual member that represents a class itself is introduced. The class-level graph is made by lifting up the member-level graph. Classes without any dependencies are not considered in this procedure. Since the authoritative decomposition was generated from the package structure of the system, to keep fair evaluation, package information was removed from dependency graphs preserving the identities of respective classes. Next, clustering algorithms are executed. All clustering tools are executed with the default parameters. Finally, the performances of the computed decompositions are measured using the aforementioned framework [19].

At the extracting dependency graph step, if multiple classes exist in a source file, we virtually treat all subsidiary classes, such as inner classes, subsequent classes and so on, as one class (i.e. the top-level public class), because they trivially belong to the same group for the clustering purpose.

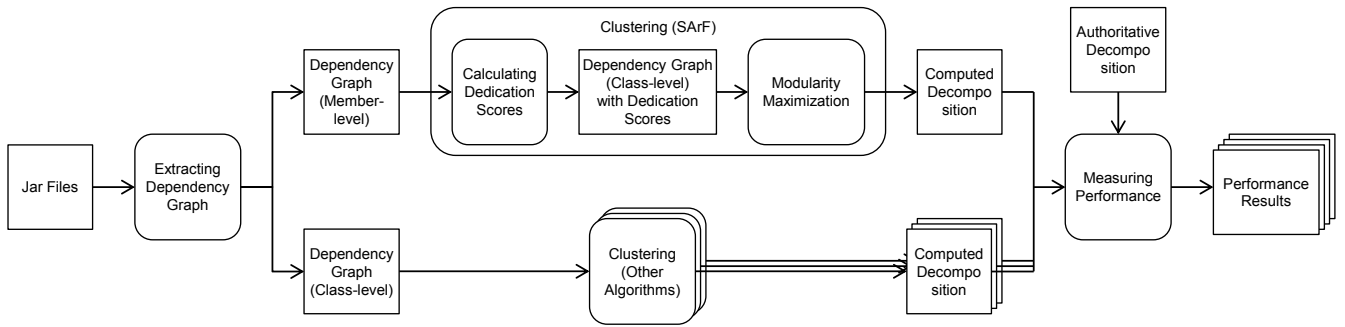

Figure 3. Evaluation Procedure

## V. CASE STUDIES

To examine SArF, we did two case studies. For each case study, the authoritative decomposition of the target software system can be obtained from a feature viewpoint. It fits the objective of SArF. The aim of the first case study is to show the concrete example results of SArF and other algorithms and to compare them. The primary aim of the second case study is to show SArF fits better to the reconstructed architecture from feature viewpoint than from package viewpoint.

### A. Case Study 1: Weka

Weka is a data mining tool and was used in previous studies [6][11]. The architecture of its version 3.0 is well documented in [32], and most of its packages correspond to its features [11].

In this case study, we used Weka version 3.0.6, and all 142 classes under package `weka` are used. Fig. 4 shows the architecture of Weka using its package diagram created using the information in [11][32] and its jar file. The architecture has three layers. Package `filter` is in the middle layer, and package `core` is in the bottom layer. The other packages are in the top layer. Each package except `core` corresponds to its feature or feature set. In Fig. 4, the vertical axis represents the level of layers, and features are aligned with the horizontal axis.

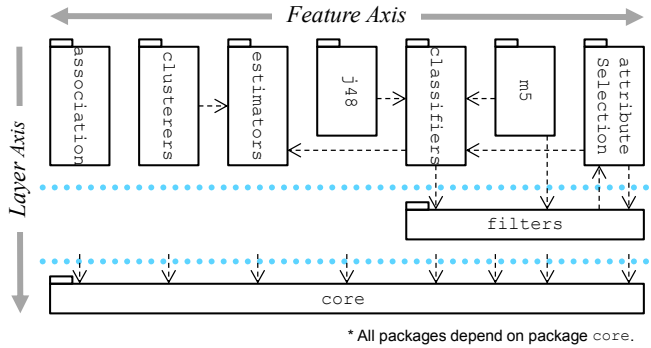

Figure 4. Architecture of Weka 3.0

We ran clustering algorithms according to Fig. 3. To assess the effectiveness of the Dedication score, we also ran the unweighted Newman algorithm using Modularity $Q_D$ on a class-level graph, and we call it "Newman" in this paper. The clustering results are shown in the first row of Table I and Fig. 5.

In Table I, column "Classes" shows the number of classes in the system, and column "$Ka$" shows the number of clusters in the authoritative decomposition. Each column labeled "$K$" shows the number of the clusters in the corresponding computed decomposition. Each column labeled "MoJoFM" shows the MoJoFM value of the decomposition. In the table, the results of algorithms, SArF, Newman, Bunch, ACDC, and the algorithm of Patel et al. are shown.

In Fig. 5, the results are visualized using *Distribution Map* [33] technique. We overlaid the technique on the architectural view in Fig. 4. Each color alphabetized small square represents a class. Its color and alphabet represents the cluster where it belongs. Groups of squares are arranged so that relevant classes in a cluster are closely located.

Fig. 5(a) shows the result of SArF. The result shows that each cluster almost matches each package except `core`. It means the algorithm effectively partitioned classes into the corresponding features, and it is consistent with the result that MoJoFM of SArF is the highest (72.9%) in Table I. Fig. 5(b) is the results of ACDC, and Fig. 5(c) is the results of Bunch. Common to both ACDC and Bunch, a few clusters widely lay on several packages through packages `filter` and `core`. It can be said that the omnipresent modules in both packages prevent the algorithms from effectively working. There are too many small clusters scattered in the ACDC result. On the other hand, multiple packages are amalgamated in the Bunch result.

The second row "Weka 3.0.6 (w/o OM)" shows the results of the subset of "Weka 3.0.6" without omnipresent modules. To remove omnipresent modules, packages `filters` and `core` are entirely removed from the dependency graph before clustering. This treatment was used by Patel et al. [11], and their result is cited in the row. The treatment was also used in [6]. In the second row, all results become higher. Especially, SArF (97.0%) and Newman (93.9%) are very high. It means unless omnipresent modules affected clustering algorithms, they could generate sufficiently proper decompositions. Another notable point is that SArF and Newman excel to the result of the dynamic feature extraction approach by Patel et al.

To normalize the comparison of the results in the second row, the results, where clustering was executed on the whole graph but the performances were measured on the selected packages excluding `filters` and `core`, are presented in the third row, "Weka 3.0.6 (selected)". Since package `core` does not correspond to a specific feature, it acts as a noise source for the purpose of measuring the effectiveness of feature-gathering clustering. The result of SArF is 90.9% and still very high. For other algorithm, the results are fairly lower than the results in the second row. It means SArF is tolerant of omnipresent modules.

More findings exist in the SArF result. Although package `filters` was removed as omnipresent modules in [11], Fig. 5(a) shows SArF could detect most of filter-relevant features (purple in Fig. 5(a)). In addition, we examined the SArF result in package `core` and found some classes are used by multiple packages but mainly by one package. It is difficult for existing omnipresent-module-removing techniques [4][8][12] to detect such classes. SArF could successfully detect them.

Comparing the first and second rows, we hypothesized as follows:

TABLE I. PERFORMANCE MEASUREMENT RESULTS ON WEKA

| System | Classes | Ka | SArF | | Newman | | ACDC | | Bunch* | | Patel et al. | |
|---|---|---|---|---|---|---|---|---|---|---|---|---|
| | | | MoJoFM | K | MoJoFM | K | MoJoFM | K | MoJoFM | K | MoJoFM | K |
| Weka 3.0.6 | 142 | 9 | **72.9** | 8 | 63.2 | 5 | 42.9 | 16 | 44.1 | 3-7 | | |
| Weka 3.0.6 (w/o OM) | 106 | 7 | **97.0** | 9 | 93.9 | 6 | 85.9 | 5 | 70.1 | 5-9 | *87.83* | 8 |
| Weka 3.0.6 (selected) | | | **90.9** | 8 | 84.9 | 5 | 55.6 | 15 | 59.6 | 3-7 | | |

In all the tables, **bold figures** represent the highest values, and *italic figures* represent the figures are cited from the original studies.
* Since Bunch is a heavily randomized algorithm, each value is the average of five measurements. The same applies to all the tables.

**Hypothesis:** The upper limit of measurable authoritativeness exists depending on the combination of the objective of a clustering algorithm and an authoritative decomposition.

For example, since featureless package `core` occupies about 20% classes, the upper limit of MoJoFM in Weka 3.0.6 seems to be 80%.

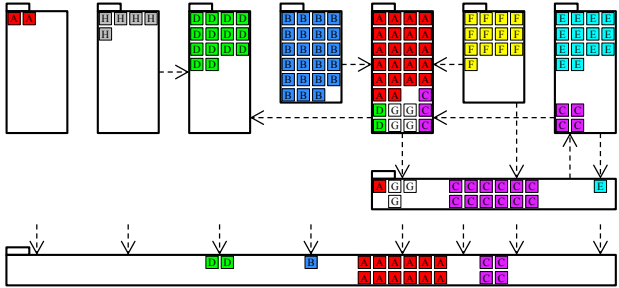

(a) Clustering Result of SArF

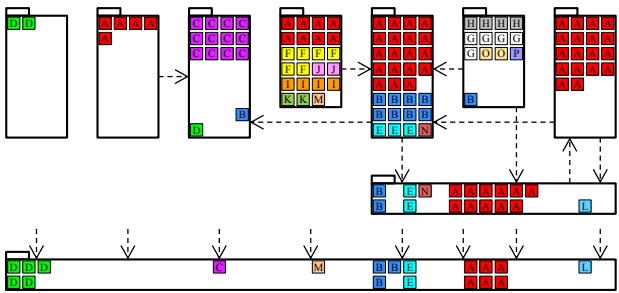

(b) Clustering Result of ACDC

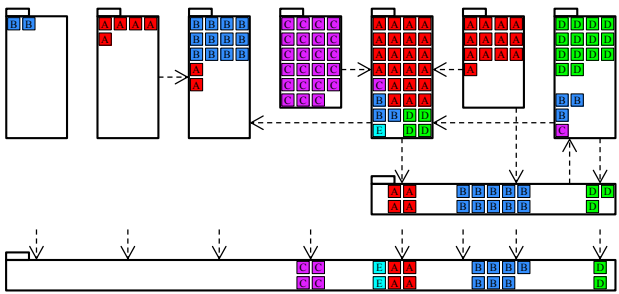

(c) Clustering Result of Bunch

Figure 5.  Visualization of Clustering Results on Weka

*B. Case Study 2: A Proprietary Data Mining Tool*

In the second case study, we used an industrial data mining product of Fujitsu tentatively called *DMTool*. We prepared two authoritative decompositions of the product. The first decomposition, *ADpackage*, is automatically generated from its package structure. The second decomposition, *ADfeature*, was obtained over a series of interviews with its developers. We requested them to create the decomposition from a feature viewpoint. The two decompositions are shown in Fig. 6. Both decompositions were laid out by the developers according to its architecture.

Fig. 6(a) shows the architecture of DMTool from a package viewpoint. Each package name is obscured for reasons of confidentiality. As shown in the figure, DMTool has two major

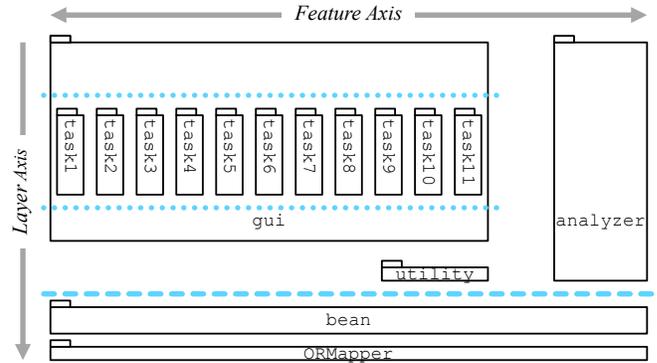

(a) Architecture from a package viewpoint

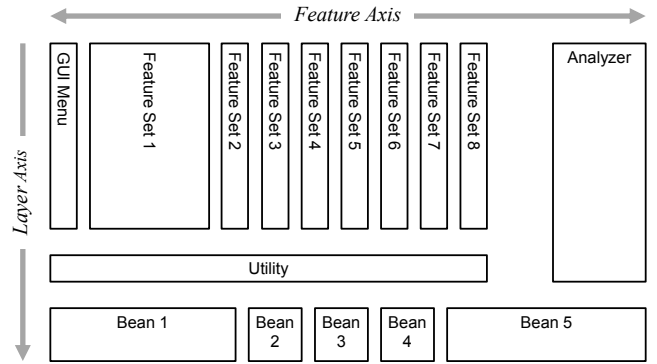

(b) Manually created architecture from a feature viewpoint

Figure 6.  Architecture of DMTool

layers, and package `gui` has three internal layers. The higher GUI layer comprises menus and widgets. The middle GUI layer comprises several user and system tasks, and some of them are implemented in nested packages shown as `task` *x*. The lower GUI layer comprises communicators to other layers.

Fig. 6(b) shows the architecture of DMTool from a feature viewpoint. In this view, package `gui` is divided into several feature set groups. Several utility classes in the lower GUI layer are packed into group Utility. In the bottom layer, bean and OR-mapper classes are rearranged into several bean groups according to their functionalities and interactions.

By comparing Fig. 6(a) and 6(b), it is found that the design policy of the package structure of DMTool is aligned to layers rather than features. It is often the case in enterprise applications. An interesting observation is that ADpackage is almost orthogonal to ADfeature, i.e., ADpackage is divided vertically, and ADfeature is divided horizontally except package `analyzer`. The difference may affect the authoritativeness results.

The clustering result of SArF is shown in Fig. 7(a) and 7(b). Both figures show the identical clustering result in the different views. The performance measurements of SArF and other algorithms are shown in Table II. Both rows in the table show the performances of the identical clustering result for each algorithm from a two viewpoints.

First, we focus on Fig. 7(b). The clustering result of SArF matches the feature-based authoritative decomposition very well except group Feature Set 1 (FS1) and group Utility. From the interviews with the developers, they put classes relevant to multiple features into FS1 or Utility, when they could not

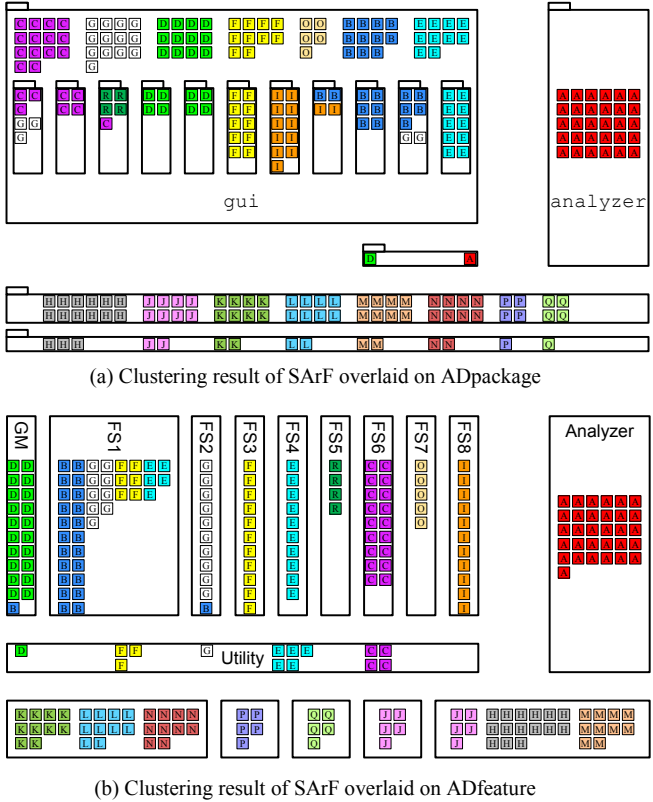

(a) Clustering result of SArF overlaid on ADpackage

(b) Clustering result of SArF overlaid on ADfeature

Figure 7. Visualization of Clustering Results on DMTool

TABLE II. PERFORMANCE MESUREMENT RESULTS ON DMTOOL

| System | Classes | Ka | SArF MoJoFM | SArF K | Newman MoJoFM | Newman K | ACDC MoJoFM | ACDC K | Bunch* MoJoFM | Bunch* K |
|---|---|---|---|---|---|---|---|---|---|---|
| DMTool (ADpackage) | 253 | 16 | 68.6 | 18 | 50.2 | 6 | 60.7 | 58 | 47.5 | 3-18 |
| DMTool (ADfeature) | | 16 | 81.4 | | 36.3 | | 65.4 | | 40.7 | |

## VI. COMPARATIVE ANALYSIS

In this section, we performed comparative evaluations and discuss the results.

### A. Target Software Systems and Compared Studies

The target software systems of the evaluations are described in Table III. The parenthesized figures after the versions mean the number of the versions. The first two systems are deemed to be packaged from a feature viewpoint and be suitable to assess the objective of SArF. The remainders were selected from the systems used in previous studies to remove biases.

Surprisingly, we found that some of the systems used in previous studies have very few packages or one of the packages occupies the majority of the system. To check such a situation, we define *Occupancy* as the percent of the classes occupied by the largest package. Package structures with high Occupancy are inappropriate as authoritative decompositions. For example, Occupancy of JFreeChart 0.9.2 used in [6][15] is 71%, and Occupancy of EasyMock 2.4 used in [16][20] is 46%, and so on. We argue systems with Occupancy less than at most 40% should be selected. The rationale of 40% is that a system almost equally divided into three packages can be accepted. The collected systems shown in Table III have less than 40% Occupancy except JabRef shown in Fig. 9.

TABLE III. DESCRIPTIVE INFORMATION OF TARGET SYSTEMS

| System (versions) | Description | Previous Studies |
|---|---|---|
| Weka 3.0.6-3.6.3 (6) | Data mining tool | [6][11] |
| Javassist 2.4-3.16.1 (19) | Byte code manipulation library | - |
| Ant 1.5-1.8.3 (17) | Software build tool | [5] |
| JUnit 4.6 | Unit testing framework | [6] |
| JHotDraw 60b1 | GUI FW for drawing editors | [15] |
| JDK Swing 1.4.0 | Java GUI widget toolkit | [16] |
| PMD 4.2.5 | Source code static analyzer | [16] |
| JabRef 1.0-2.4b (35) | Bibliography reference manager | [20] |

### B. Results

We measured the performances of the algorithms, SArF, Newman, ACDC, and Bunch for the respective target systems according to the procedure in Fig. 3. The results are shown in Table IV and V. In Table IV, the right four columns show the NED values for the respective algorithms. For the first three systems, the averages of the series of the measured stability values are shown in Table V.

In respect of authoritativeness, in Table IV, SArF has the highest MoJoFM value in the every system. In most systems, ACDC is the second, Newman is the third, and Bunch is the last. Since Javassist has a feature-based package structure, it is reasonable that MoJoFM of SArF on Javassist is much higher (63-76%) than other algorithms. Since the later versions of Weka have non-feature-based packages, MoJoFM of SArF on Weka become poorer in the later version (from 73% to 48%) but is still the best.

determine one group to be put into. Such cases often happen, because it is a well-observed developer's habit [4].

In the second row of Table II is the MoJoFM values measured on ADfeature. MoJoFM of SArF is 81.4%, high and the highest, and it is consistent with the above observation.

Then, we go back to observe Fig. 7(a). In the figure, only some parts of the clustering result matches the packages. Some clusters in the SArF result are scattered in package `gui`, `bean`, and `ORmapper`. In the first row of Table II is the MoJoFM values measured on ADpackage. MoJoFM of SArF is 68.6% and still the highest; however it is much poorer than the value on ADfeature. It means that ADfeature is appropriate for SArF to measure its effectiveness and that ADpackage is not so appropriate. Since SArF achieved high MoJoFM on ADfeature, it can be said SArF works well to gather features in the case.

The differences between SArF and Newman are greater than in Weka. It is caused by the susceptibleness to omnipresent modules. MoJoFM of Newman on ADfeature is poorer than on ADpackage. The reason is that Newman tends to generate large few clusters (refer to column "*K*") and there is no large group such as package `gui` in ADfeature. About this topic, we will discuss further in Section VI.C.

To estimate the upper limit of authoritativeness of SArF on ADpackage, assuming that the ideal of SArF is ADfeature, we measured MoJoFM(ADfeature, ADpackage), and it is 73.6%. Since MoJoFM of SArF on ADpackage is 68.6%, it is close to the limit. As long as ADpackage is used as a benchmark, it can be said that feature-gathering clustering cannot be improved.

Finally, to confirm the validity of the clustering result of SArF, we consulted the developers again. They confirmed the result was acceptable as one of feature views of DMTool.

TABLE IV. AUTHORITATIVENESS AND NON-EXTREMITY OF CLUSTER DISTRIBUTION (NED) RESULTS

| System (versions) | Classes | Ka | Authoritativeness | | | | | | | | NED (average) | | | |
|---|---|---|---|---|---|---|---|---|---|---|---|---|---|---|
| | | | SArF | | Newman | | ACDC | | Bunch* | | SArF | Newman | ACDC | Bunch* |
| | | | MoJoFM | K | MoJoFM | K | MoJoFM | K | MoJoFM | K | | | | |
| Weka 3.0.6-3.6.3 (6) | 142-1114 | 9-56 | **48-73** | 8-34 | 31-63 | 5-17 | 45-55 | 16-150 | 17-41 | 3-48 | **0.944** | 0.563 | 0.636 | 0.538 |
| Javassist 2.4-3.16.1 (19) | 121-206 | 8-13 | **63-76** | 11-15 | 42-59 | 5-7 | 53-63 | 17-29 | 30-48 | 3-26 | **0.710** | 0.268 | 0.457 | 0.549 |
| Ant 1.5-1.8.3 (17) | 266-694 | 10-32 | **46-58** | 21-38 | 28-50 | 7-12 | 42-53 | 27-81 | 22-47 | 4-49 | **0.916** | 0.574 | 0.604 | 0.641 |
| JUnit 4.6 | 145 | 16 | **48.5** | 11 | 36.9 | 7 | 45.1 | 23 | 26.5 | 5-7 | **0.972** | 0.276 | 0.745 | 0.465 |
| JHotDraw 60b1 | 285 | 12 | **44.7** | 17 | 34.4 | 7 | 43.2 | 42 | 24.3 | 5-16 | **0.947** | 0.077 | 0.565 | 0.530 |
| JDK Swing 1.4.0 | 536 | 15 | **46.3** | 21 | 37.6 | 6 | 42.2 | 51 | 33.0 | 8-17 | **0.981** | 0.278 | 0.884 | 0.790 |
| PMD 4.2.5 | 565 | 26 | **54.0** | 17 | 48.8 | 8 | 53.0 | 52 | 36.6 | 10-32 | 0.662 | 0.248 | 0.535 | **0.710** |

In respect of NED, SArF also has the highest NED value in the every system except PMD. It is interesting that the numbers of clusters ($K$) of SArF are close to the number of the clusters on authoritative decompositions. The facts suggest that the cluster distribution of SArF is good. Conversely, $K$ of Newman is too small, and $K$ of ACDC is too large. ACDC tends to generate a few large clusters and many tiny clusters as described in [19].

In respect of stability, in Table V, SArF is the best; however, there are little differences except Bunch. We observed the perturbation of the stability of Newman tends to be proportionally larger than SArF. It implies that the impact of unimportant dependencies is suppressed by Dedication used in SArF and that SArF is robust to the unimportant changes between versions.

TABLE V. STABILITY RESULTS

| System (versions) | SArF | Newman | ACDC | Bunch |
|---|---|---|---|---|
| Weka 3.0.6-3.6.6 (6) | **80.6** | 79.7 | 76.8 | 44.8 |
| Javassist 2.4.0-3.16.1 (19) | **96.3** | 94.9 | **96.3** | 58.7 |
| Ant 1.5-1.8.3 (17) | **94.4** | 89.3 | 83.2 | 47.7 |

*C. Discussions*

*1) SArF vs Newman algorithm*

Since the Newman algorithm is the basis of the SArF algorithm, we discuss the performance difference of the two algorithms. The aforementioned results show SArF is always obviously superior to Newman in authoritativeness and NED and is always slightly superior to Newman in stability.

The Newman algorithm was first utilized by Erdemir et al. [6] in software clustering. Since they did not describe the objective function, we could not reproduce their study. Table VI shows parts of the aforementioned results using MoJoSim and also shows the results of Erdemir et al. The first two systems in the table were measured in the same conditions as [6]. Since the underlined results are very similar, we infer the two utilizations are almost the same.

Table VI also presents the symptoms of the flaw of MoJoSim pointed out by Wen and Tzerpos [23]. In the every system, MoJoSim of Newman is higher than MoJoSim of SArF, and on the contrary, MoJoFM of Newman is lower than MoJoFM of SArF. The reason of the contrariety is explainable. As shown in columns "$K$", the clusters generated by Newman tend to be fewer than the clusters of the authoritative decomposition. The tendency is more eminent in larger systems. Newman can be said to be susceptible to omnipresent modules. The same was observed in the case study 2. As previously explained, since MoJoSim overestimates a decomposition with a few large clusters, MoJoSim of Newman tends to be high, and the contrariety occurred.

Even if the number of clusters is too small, the result of hierarchical clustering has a chance to be manually expanded to the appropriate number of clusters. However, in the case of Newman, such a manual expansion does not work well. Since the algorithm is greedy, the obtained merge tree tends to be strongly unbalanced, and it may be difficult to find useful cut points as shown in Fig. 8(a). Fig. 8 shows the merge trees generated by Newman and SArF from "Weka 3.0.6". On the contrary, the merge trees generated by SArF tend to be more balanced and dividable as shown in Fig. 8(b).

TABLE VI. RESULTS COMPARED BETWEEN SArF AND NEWMAN

| System | Ka | SArF | | | Newman | | | Erdemir |
|---|---|---|---|---|---|---|---|---|
| | | MoJoFM/Sim | | K | MoJoFM/Sim | | K | MoJoSim |
| Weka 3.0.6 (w/o OM) | 7 | 97.0 | 97.2 | 9 | 93.9 | <u>98.1</u> | 6 | <u>97.98</u> |
| JUnit 4.6 | 16 | 48.5 | 64.8 | 11 | 36.9 | <u>73.1</u> | 7 | <u>74.55</u> |
| JHotDraw 60b1 | 12 | 44.7 | 47.0 | 17 | 34.4 | 53.0 | 7 | - |
| JDK Swing 1.4.0 | 15 | 46.3 | 47.8 | 21 | 37.6 | 73.9 | 6 | - |
| PMD 4.2.5 | 26 | 54.0 | 72.7 | 17 | 48.8 | 77.2 | 8 | - |

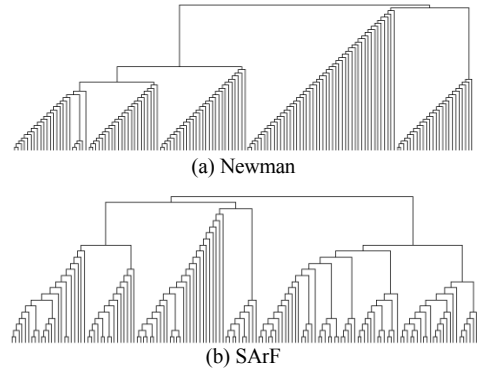

(a) Newman

(b) SArF

Figure 8. Clustering Processes (merge trees/dendrograms) on Weka

*2) Comparison with Other Graph Clustering Approaches*

Fig. 9 shows the comparison between existing software clustering approaches using graph clustering algorithms and SArF. The original chart of the figure was presented in Bittencourt and Guerrero [20] and showed the performance results of four algorithms, k-means, Bunch, GN, and design structure matrix (DSM) clustering [34], by clustering 35 versions of JabRef. In the figure, we overlaid the curve of the result of SArF with a black solid line. It shows the result of SArF is the highest in 34 of 35 versions. We can say SArF is superior to the other graph clustering approaches in this case.

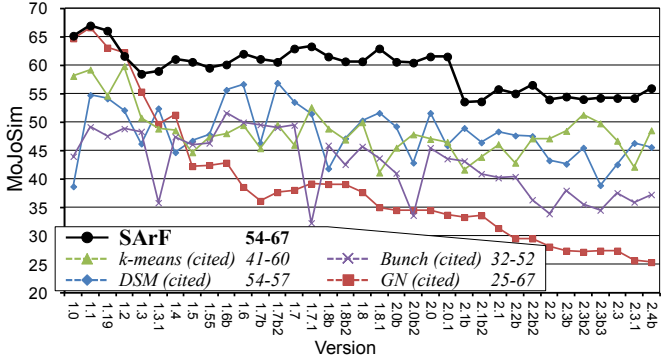

Figure 9. JabRef results compared with other graph clustering approaches

### 3) Validating the definition of SArF

Since we built many assumptions in the definition of the Dedication score in Section III.A, its validity should be assessed empirically. The following results support the validity of the definition of the Dedication score: (1) SArF outperforms Bunch with the omnipresent-module removal shown in the second row in Table I; (2) SArF always outperforms Newman; and (3) Absolute high performance achieved by SArF.

### 4) Comparison with Semantic Approaches

Table VII shows the comparison of SArF with other approaches. The upper half of the table shows the comparison with the result reported by Scanniello et al. [15]. Their approach uses k-means clustering using semantic information and is characterized by detecting architectural layers. In the table, the result of SArF is much poorer than their result. It is explainable on the basis of the objective of clustering algorithms. The objective of their approach is to detect layers, and they chose JHotDraw because of its layer-style architecture. As described in the case studies, the objective of SArF is to gather features and is orthogonal to detecting layers. Therefore, it can be said that JHotDraw is suited for their approach, but not so suited for SArF. It is to be noted that the result of SArF is not poor as shown in the row of JHotDraw in Table IV.

The lower half of Table VII shows the comparison with the results in Corazza et al. [16]. Their approach uses hierarchical clustering using semantic information and is characterized by automatically weighting different lexical zones using EM algorithm. The results in the table show that the performance behaviors of their approach and SArF are very different. Swing is clustered well by their approach but not by SArF. PMD is clustered well by SArF but not by their approach. Since the objective of their approach is not definitely defined, we cannot infer the reasons of the difference as the previous case.

TABLE VII. RESULTS COMPARED WITH SEMANTIC APPROACHES

| Approach:<br>Scanniello et al. [15] | System | SArF<br>MoJoSim | Scanniello<br>MoJoSim |
|---|---|---|---|
| Layer detection and semantic-based clustering | JHotDraw 60b1 | 47.0 | **86** |
| **Approach:<br>Corazza et al. [16]** | **System** | **SArF<br>MoJoSim** | **Corazza<br>MoJoSim** |
| Semantic-based clustering with auto-weighted lexical zones | JDK Swing 1.4.0 | 47.8 | **67.5** |
|  | PMD 4.2.5 | **72.7** | 47.0 |

### 5) On Authoritative Decomposition

In the case study one, we made the hypothesis, "The upper limit of measurable authoritativeness exists." In both case studies, the hypothesis was supported. The very high MoJoFM of SArF on Javassist also supports it. Therefore, it is reasonable to support the hypothesis is true. Besides, from the comparisons with two approaches in Table VII, the followings can be said:

1. To evaluate software clustering algorithms, it is important to choose systems with appropriate authoritative decompositions suited for the objective of the algorithms.
2. If two algorithms have different characteristics, they cannot be compared. This was also pointed out by Shtern and Tzerpos [26].

## VII. THREATS TO VALIDITY

The objective of SArF is to gather features; however, since an approach without any semantic or dynamic information is novel, it needs many case studies and experiments to validate the algorithm works properly. Unfortunately, few of open source systems we examined take a feature-based packaging policy, and we found only Weka 3.0 and Javassist. As shown in our case studies, measured authoritativeness can heavily vary depending on used authoritative decompositions. Therefore, to evaluate authoritativeness precisely, the objective of the measured algorithm should be clarified and the authoritative decomposition which fits to its objective should be obtained.

Our clustering algorithm uses static dependency graphs. Static and dependency-based approach has some drawbacks such that dynamic invocations cannot be fully tracked. To overcome the problem, hybrid approaches with semantic or dynamic techniques are promising such as the approach of Scanniello et al. [15].

Another threat is that the set of performance measures seems not to be complete. We used module-based measures such as MoJoFM. Since edges are weighted in SArF, we could not employ existing edge-based measures such as EdgeSim [25]. We expect measures for hierarchical decomposition such as UpMoJo [24] fit better to practical situations. We did not use it because of incompatibility with the existing evaluations.

## VIII. CONCLUSIONS

We have proposed a novel approach of static dependency-based software clustering, SArF. It has two key characteristics. The first is that it eliminates the process of removing omnipresent modules which requires human interactions. By this elimination, software clustering process can be further automated. The second characteristic is that the objective of the algorithm is gathering modules from a feature viewpoint.

The SArF algorithm comprises of two ideas, the definition of the Dedication score and the utilization of Modularity Maximization. Dedication means the importance of a dependency for its dependent and is defined on the basis of fan-in analysis. By using Dedication, important dependencies count, and omnipresent modules are viewed as unimportant but still considered. Utilization of Modularity Maximization effectively enables finding the structures of the dependencies with Dedication.

In the two case studies, we evaluated SArF using authoritative decompositions from a feature viewpoint. The results show the decompositions computed by SArF fit to the feature-based decompositions. In addition, in spite of the omnipresent modules, SArF could decompose the two systems automatically with high authoritativeness. We also evaluated authoritativeness, NED and stability of SArF using nine systems with dozens of versions. The results show the performance of SArF is superior to existing dependency-based algorithms.

The contributions of our study are summarized as follows:

1. Features were successfully gathered only using static dependency information.
2. The omnipresent-module-removing step is eliminated in dependency-based software clustering.
3. The extensive performance evaluations show that SArF is superior to existing dependency-based software clustering algorithms in various criteria.
4. The existence of the upper limit of measurable authoritativeness and the necessity of collecting suitable authoritative decomposition for the objective of the evaluated software clustering algorithm were pointed out.

In Wu et al. [19], since their authoritativeness results were low, they described the used algorithms might be not mature. However, the existence of the upper limit may break their negative consideration. Since some recent algorithms such as [15] and SArF scored high (over 70%) MoJoSim or MoJoFM values in appropriate authoritative decompositions, we conclude software clustering is promising.

As future work, to cover some drawbacks of static dependency-based approaches, we plan to examine hybrid approaches using semantic and dynamic approaches. Since the concept of Dedication is not restricted in static dependency, other information such as co-change information and developer networks can be involved. We also plan to combine techniques in source code summarization and feature location with SArF. Finally, without appropriate benchmarks, it is difficult to improve the accuracy of algorithms. Therefore, more practical performance measures or measurement methods are required.